\documentclass[a4paper,11pt]{article}
\usepackage{pos}

\newcommand{\meV}{\,\mathrm{meV}}

\newcommand{\cmsq}{\,\mathrm{cm^2}}

\newcommand{\nchi}{n_\chi}
\newcommand{\bq}{\boldsymbol{q}}
\newcommand{\oq}{\omega_{\boldsymbol{q}}}
\newcommand{\bv}{\boldsymbol{v}}
\newcommand{\fmed}{F_\mathrm{med}}
\newcommand{\sigman}{\sigma_{\chi N}}

\title{Detecting Dark Matter Induced Power in Quantum Devices}

\author*[a]{Anirban Das}
\author[b,c]{Noah Kurinsky,}
\author[b,c]{Rebecca K. Leane}

\affiliation[a]{Department of Physics \& Astronomy, Seoul National University,
  Seoul 08826, South Korea}

\affiliation[b]{SLAC National Accelerator Laboratory, 2575 Sand Hill Rd, Menlo Park, CA 94025, USA}

\affiliation[c]{Kavli Institute for Particle Astrophysics and Cosmology, Stanford University, Stanford, CA 94035, USA}

\emailAdd{anirbandas@snu.ac.kr}
\emailAdd{kurinsky@slac.stanford.edu}
\emailAdd{rleane@slac.stanford.edu}

\abstract{In the past few years, many mesoscale systems have been proposed
as possible detectors of sub-GeV dark matter particles. In this work, we point out the feasibility of probing dark matter-nucleon scattering cross section using superconductor-based quantum devices with meV-scale energy threshold. We compute new limits on spin-independent dark matter scattering cross section using the existing power measurement data from three different experiments for MeV to 10 GeV mass. We derive the limits for both halo and thermalized dark matter populations.}

\FullConference{XVIII International Conference on Topics in Astroparticle and Underground Physics (TAUP2023)\\
 28.08-01.09.2023\\
University of Vienna\\}

\begin{document}
\maketitle

\section{Introduction}
Low energy threshold devices have heralded a new era in the search for light dark matter (DM). The availability of mesoscale devices based on superconductors, such as transition edge sensor, microwave kinetic inductance detector etc. enable us detect small $\lesssim\mathcal{O}(100)$\,meV energy depositions that can arise from sub-MeV DM scattering. Moreover, today's quantum devices can measure very small amounts of power deposition in the form of quasiparticle excitations from Cooper pair breaking in superconductors. Since the typical binding energy of a Cooper pair is $\lesssim 1\,\mathrm{meV}$, these devices have sensitivity to very low energy as they can detect even single quasiparticle. In Ref.\,\cite{Das:2022srn}, we showed using the data from existing experiments that we can probe DM-nuclear scattering cross sections down to $\sim 10^{-29}\cmsq$ for GeV-mass halo DM. 

In some DM theories where it interacts relatively strongly with the ordinary matter, it is possible for DM to get captured inside the earth. This capture process builds up a thermalized population of DM that can have orders of magnitude larger number density than the halo population\,\cite{Neufeld:2018slx,Pospelov:2020ktu, Pospelov:2019vuf, Xu:2021lmg,Budker:2021quh, McKeen:2022poo,Billard:2022cqd, Leane:2022hkk,Acevedo:2023xnu}. However, this thermalized population will have the local ordinary matter temperature of roughly $300$\,K, or equivalently, $\sim 10\meV$ energy. Therefore, meV-threshold devices are needed to probe this captured/thermalized DM too.

\section{Scattering Rate \& Detection Mechanism}
We only consider scattering between DM and the nucleus of the detector material which will be aluminum (Al) and silicon (Si). We follow Ref.\,\cite{Campbell-Deem:2022fqm} to compute the energy deposition through single and multiphonon excitation from scattering.
\begin{align}\label{eq:rate}
    \Gamma = \frac{\pi\sigman\nchi}{\rho_T\mu^2} \int d^3v f_\chi(\bv) \int \frac{d^3q}{(2\pi)^3} \fmed^2(q) S(\bq,\oq)
\end{align}
Here, $f_\chi(\bv)$ is the Maxwellian DM velocity distribution, $\rho_T$ is the target material density, $\sigman$ is the DM-nucleon scattering cross section, $\mu$ is the reduced mass of the DM-nucleon pair, $\fmed(q)$ is a form factor that depends on the mediator mass, and $S(\bq,\oq)$ is the dynamic structure factor that encodes the detector response to DM scattering which we compute analytically. We use the public code DarkELF with appropriate modifications for the thermalized DM parameters and the detector crystal structure\,\cite{Campbell-Deem:2022fqm}. The power injected from DM scattering is computed as follows,
\begin{align}\label{eq:P_DM}
    P_\mathrm{DM} = \epsilon\int d\omega~\omega\dfrac{d\Gamma}{d\omega}\,,
\end{align}
where $\epsilon$ is an efficiency factor. To calculate the quasiparticle generation rate in Al, we simply divide Eq.(\ref{eq:P_DM}) by the Cooper pair binding energy $\Delta=0.34\meV$. 

\section{Experimental Data}
In our work Ref.\,\cite{Das:2022srn}, we used three different experiments to constrain spin-independent DM-nucleon scattering cross section.

\emph{Low Quasiparticle Background Devices:} Ref.\,\cite{Mannila_2021} performed an experiment to test how low the quasiparticle background can be in an Al superconducting island. They monitored the quasiparticle density on the island and found a value $0.013\,\mathrm{\mu m^{-3}}$. We converted this quasiparticle density into a power density $P=6\times10^{-25}\,\mathrm{W\mu m^{-3}}$ which we compare with the DM-injected power density.
\begin{figure}[t]
    \centering
    \includegraphics[width=0.47\textwidth]{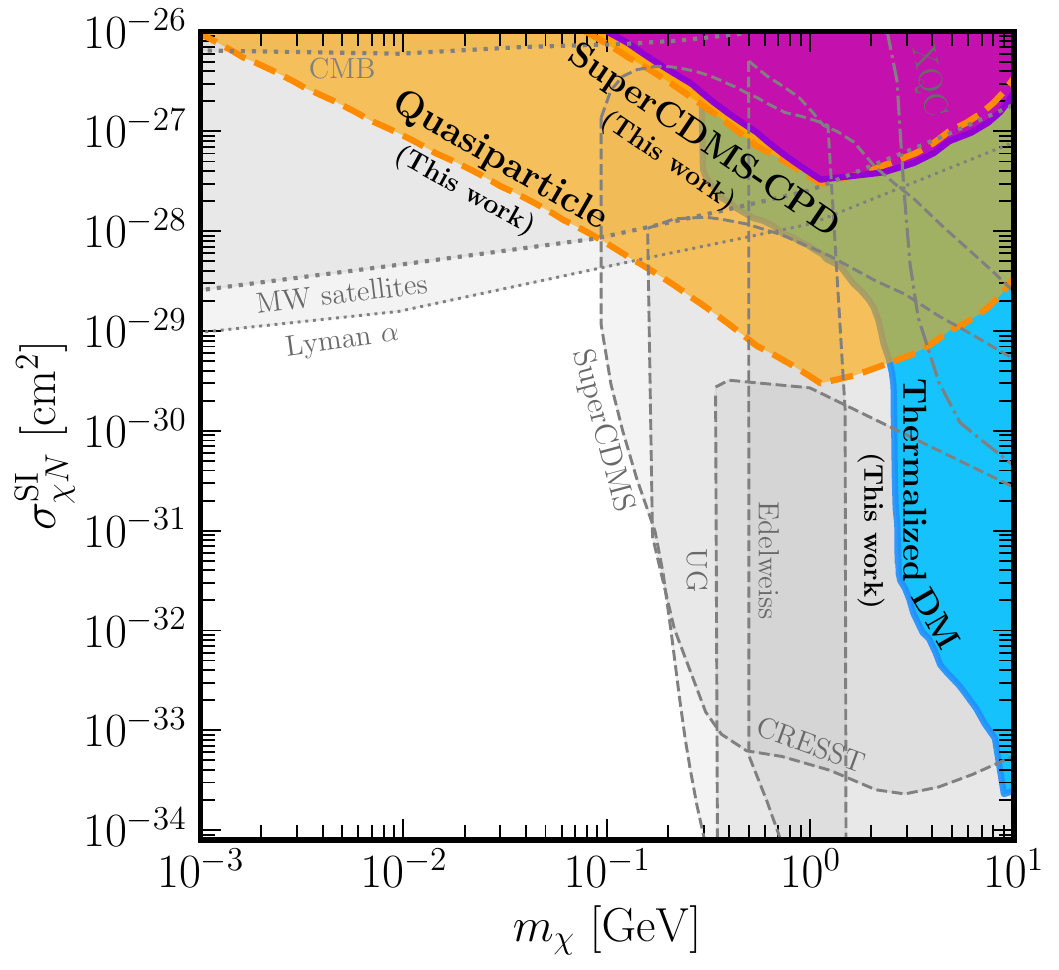}
    \caption{New limits on spin-independent DM-nucleon scattering cross section from Ref.\,\cite{Das:2022srn}. We show the halo DM limits from excess quasiparticle production measurements (orange) and SuperCDMS-CPD
(magenta). The blue-shaded region is the new constraint
on the thermalized DM population. The gray-shaded regions are already excluded by other laboratory experiments and astrophysical observations.}
    \label{fig:limit_plot}
\end{figure}

\emph{Low Noise Bolomters:} In Ref.\,\cite{QCD}, the authors developed a single photon detector with superconducting Al. They used a quantum capacitance device to detect the photon-produced electrons. In the absence of any external source, their device measured an excess power of $P=4\times10^{-20}\,\mathrm{W\mu m^{-3}}$ which we used in our work.

\emph{SuperCDMS Dark Matter Detector:} SuperCDMS-CPD is a conventional DM direct detection experiment that used a 10.6 g Si block as the absorber material\,\cite{Fink_2021}. In their experimental run, they measured an excess power 18 pW in the Si. This translates to a power density $10^{-24}\,\mathrm{W\mu m^{-3}}$.

\section{Results \& Discussion}
The main results of this work are shown in Fig.\,\ref{fig:limit_plot}. The strongest bound is obtained from the \emph{low quasiparticle background} experiment (orange) which reaches below $\sigman\simeq10^{-29}\cmsq$ around 1 GeV DM mass. This is followed by the limit from the SuperCDMS-CPD experiment (magenta). The blue region in Fig.\,\ref{fig:limit_plot} shows the constraints for the thermalized DM. In this case, all three experimental limits overlap and are limited by the evaporation of lighter DM. The bounds from the other laboratory experiments and astrophysical observations are shown in gray. Even though the new limits overlap with other existing constraints, they are subject to improve in future with more dedicated experiments. We do not show any ceiling for our new limits as these experiments are all surface-run and the halo DM would still be present for the largest scattering cross section shown here. 
This work adds a qualitatively new direction in the DM direct detection experiments and demonstrates the potential of using mesoscopic quantum devices for DM search. This will inspire more future works to study and better understand the quasiparticle generation process, and other systematics in these experiments which is essential to claim any future discovery.

\bibliographystyle{jhep}
\bibliography{DM_noise}

\providecommand{\href}[2]{#2}\begingroup\raggedright\begin{thebibliography}{10}

\bibitem{Das:2022srn}
A.~Das, N.~Kurinsky, and R.~K. Leane, {\it {Dark Matter Induced Power in Quantum Devices}},  \href{http://arxiv.org/abs/2210.09313}{{\tt arXiv:2210.09313}}.

\bibitem{Neufeld:2018slx}
D.~A. Neufeld, G.~R. Farrar, and C.~F. McKee, {\it {Dark Matter that Interacts with Baryons: Density Distribution within the Earth and New Constraints on the Interaction Cross-section}},  {\em Astrophys. J.} {\bf 866} (2018), no.~2 111, [\href{http://arxiv.org/abs/1805.08794}{{\tt arXiv:1805.08794}}].

\bibitem{Pospelov:2020ktu}
M.~Pospelov and H.~Ramani, {\it {Earth-bound millicharge relics}},  {\em Phys. Rev. D} {\bf 103} (2021), no.~11 115031, [\href{http://arxiv.org/abs/2012.03957}{{\tt arXiv:2012.03957}}].

\bibitem{Pospelov:2019vuf}
M.~Pospelov, S.~Rajendran, and H.~Ramani, {\it {Metastable Nuclear Isomers as Dark Matter Accelerators}},  {\em Phys. Rev. D} {\bf 101} (2020), no.~5 055001, [\href{http://arxiv.org/abs/1907.00011}{{\tt arXiv:1907.00011}}].

\bibitem{Xu:2021lmg}
X.~Xu and G.~R. Farrar, {\it {Constraints on GeV Dark Matter interaction with baryons, from a novel Dewar experiment}},  \href{http://arxiv.org/abs/2112.00707}{{\tt arXiv:2112.00707}}.

\bibitem{Budker:2021quh}
D.~Budker, P.~W. Graham, H.~Ramani, F.~Schmidt-Kaler, C.~Smorra, and S.~Ulmer, {\it {Millicharged Dark Matter Detection with Ion Traps}},  {\em PRX Quantum} {\bf 3} (2022), no.~1 010330, [\href{http://arxiv.org/abs/2108.05283}{{\tt arXiv:2108.05283}}].

\bibitem{McKeen:2022poo}
D.~McKeen, M.~Moore, D.~E. Morrissey, M.~Pospelov, and H.~Ramani, {\it {Accelerating Earth-Bound Dark Matter}},  \href{http://arxiv.org/abs/2202.08840}{{\tt arXiv:2202.08840}}.

\bibitem{Billard:2022cqd}
J.~Billard, M.~Pyle, S.~Rajendran, and H.~Ramani, {\it {Calorimetric Detection of Dark Matter}},  \href{http://arxiv.org/abs/2208.05485}{{\tt arXiv:2208.05485}}.

\bibitem{Leane:2022hkk}
R.~K. Leane and J.~Smirnov, {\it {Floating Dark Matter in Celestial Bodies}},  \href{http://arxiv.org/abs/2209.09834}{{\tt arXiv:2209.09834}}.

\bibitem{Acevedo:2023xnu}
J.~F. Acevedo, R.~K. Leane, and L.~Santos-Olmsted, {\it {Milky Way White Dwarfs as Sub-GeV to Multi-TeV Dark Matter Detectors}},  \href{http://arxiv.org/abs/2309.10843}{{\tt arXiv:2309.10843}}.

\bibitem{Campbell-Deem:2022fqm}
B.~Campbell-Deem, S.~Knapen, T.~Lin, and E.~Villarama, {\it {Dark matter direct detection from the single phonon to the nuclear recoil regime}},  \href{http://arxiv.org/abs/2205.02250}{{\tt arXiv:2205.02250}}.

\bibitem{Mannila_2021}
E.~T. Mannila, P.~Samuelsson, S.~Simbierowicz, J.~T. Peltonen, V.~Vesterinen, L.~GrÃ¶nberg, J.~Hassel, V.~F. Maisi, and J.~P. Pekola, {\it A superconductor free of quasiparticles for seconds},  {\em Nature Physics} {\bf 18} (dec, 2021) 145--148.

\bibitem{QCD}
P.~M. {Echternach}, B.~J. {Pepper}, T.~{Reck}, and C.~M. {Bradford}, {\it {Single photon detection of 1.5 THz radiation with the quantum capacitance detector}},  {\em Nature Astronomy} {\bf 2} (Nov., 2018) 90--97.

\bibitem{Fink_2021}
C.~W. Fink, S.~L. Watkins, T.~Aramaki, P.~L. Brink, J.~Camilleri, X.~Defay, S.~Ganjam, Y.~G. Kolomensky, R.~Mahapatra, N.~Mirabolfathi, W.~A. Page, R.~Partridge, M.~Platt, M.~Pyle, B.~Sadoulet, B.~Serfass, and S.~Z. and, {\it Performance of a large area photon detector for rare event search applications},  {\em Applied Physics Letters} {\bf 118} (jan, 2021) 022601.

\end{thebibliography}\endgroup



\end{document}